\pdfoutput=1
\documentclass[a4paper,american,citeautoscript,floatfix,pdftex,superscriptaddress,twoside,%
aip,jcp,%
reprint,twocolumn,
]{revtex4-1}%
\usepackage{amsfonts,amsmath,amssymb}
\usepackage{commath}
\usepackage[T1]{fontenc}
\usepackage{graphicx}%
\usepackage[utf8]{inputenc}
\usepackage{microtype}
\usepackage[obeyFinal,textsize=footnotesize]{todonotes}
\usepackage{xspace}
\usepackage{hyperref, hypernat}
\usepackage[todos]{CMImacros}

\newcommand{\dii}{2,5-diiodothiophene\xspace}%
\newcommand{\diis}{2,5-diiodothiophene\xspace}%
\newcommand{\dqn}{\ensuremath{\Delta\text{q}_\text{norm}}\xspace}%

\newcommand{\cfeldesy}{\affiliation{Center for Free-Electron Laser Science, Deutsches
      Elektronen-Synchrotron DESY, Notkestraße 85, 22607 Hamburg, Germany}}%
\newcommand{\uhhchem}{\affiliation{Department of Chemistry, Universität Hamburg,
      Martin-Luther-King-Platz 6, 20146 Hamburg, Germany}}%
\newcommand{\uhhcui}{\affiliation{Center for Ultrafast Imaging, Universität Hamburg, Luruper
      Chaussee 149, 22761 Hamburg, Germany}}%
\newcommand{\uhhphys}{\affiliation{Department of Physics, Universität Hamburg, Luruper Chaussee 149,
      22761 Hamburg, Germany}}%
\newcommand{\unibas}{\altaffiliation[Present address:~]{Department of Chemistry, University of
      Basel, Klingelbergstrasse 80, 4056, Basel, Switzerland}}%
\newcommand{\melbourne}{\altaffiliation[Present address:~]{ARC Centre in Advanced Molecular Imaging,
      School of Physics, The University of Melbourne, Parkville 3010, Australia}}%
\newcommand{\lcls}{\affiliation{LCLS, SLAC National Accelerator Laboratory, Menlo Park, CA, 94025,
      USA}}%
\newcommand{\xfel}{\affiliation{European XFEL GmbH, 22869 Schenefeld, Germany}}%
\newcommand{\altxfel}{\altaffiliation[Present address:~]{European XFEL GmbH, 22869 Schenefeld,
      Germany}}%
\newcommand{\desy}{\affiliation{Deutsches Elektronen-Synchrotron DESY, 22607 Hamburg, Germany}}
\newcommand{\pulse}{\affiliation{SLAC National Accelerator Laboratory, PULSE Institute, Stanford,
      CA, 94305, USA}}%
\newcommand{\stanuni}{\affiliation{Department of Physics, Stanford University, Stanford, CA 94305,
      USA}}%
\newcommand{\uniaar}{\affiliation{Department of Chemistry, Aarhus University, 8000 Aarhus C,
      Denmark}}%
\newcommand{\ksu}{\affiliation{J.~R.\ Macdonald Laboratory, Department of Physics, Kansas State
      University, Manhatten, KS, 66506, USA}}%
\newcommand{\mpghd}{\affiliation{Max Planck Institute for Nuclear Physics, 69117 Heidelberg,
      Germany}}%

\newcommand{\jkemail}{\email[Corresponding author. Email:~]{jochen.kuepper@cfel.de}}%
\newcommand{\cmiweb}{\homepage[website:~]{https://www.controlled-molecule-imaging.org}}%

\begin{document}
\title{X-ray diffractive imaging of controlled gas-phase molecules: Toward imaging of dynamics in
   the molecular frame}%
\author{Thomas Kierspel}\unibas\cfeldesy\uhhcui\uhhphys%
\author{Andrew Morgan}\melbourne\cfeldesy%
\author{Joss Wiese}\cfeldesy\uhhchem%
\author{Terry Mullins}\cfeldesy%
\author{Andy Aquila}\lcls%
\author{Anton Barty}\cfeldesy%
\author{Richard Bean}\cfeldesy\xfel%
\author{Rebecca Boll}\altxfel\desy%
\author{S\'{e}bastien Boutet}\lcls%
\author{Philip Bucksbaum}\lcls\pulse\stanuni%
\author{Henry~N.~Chapman}\cfeldesy\uhhcui\uhhphys%
\author{Lauge Christensen}\uniaar%
\author{Alan~Fry}\lcls\pulse%
\author{Mark~Hunter}\lcls%
\author{Jason~E.~Koglin}\lcls%
\author{Mengning Liang}\lcls%
\author{Valerio Mariani}\cfeldesy%
\author{Adi Natan}\pulse%
\author{Joseph Robinson}\lcls%
\author{Daniel Rolles}\desy\ksu%
\author{Artem Rudenko}\ksu%
\author{Kirsten Schnorr}\mpghd%
\author{Henrik Stapelfeldt}\uniaar%
\author{Stephan Stern}\cfeldesy%
\author{Jan Th{\o}gersen}\uniaar%
\author{Chun Hong Yoon}\cfeldesy\xfel%
\author{Fenglin Wang}\cfeldesy\pulse%
\author{Jochen~Küpper}\jkemail\cmiweb\cfeldesy\uhhcui\uhhphys\uhhchem%
\date{\today}
\begin{abstract}
   We report experimental results on the diffractive imaging of three-dimensionally aligned \dii
   molecules. The molecules were aligned by chirped near-infrared laser pulses, and their structure
   was probed at a photon energy of 9.5~keV ($\lambda\approx130~\text{pm}$) provided by the Linac
   Coherent Light Source. Diffracted photons were recorded on the CSPAD detector and a
   two-dimensional diffraction pattern of the equilibrium structure of \dii was recorded. The
   retrieved distance between the two iodine atoms agrees with the quantum-chemically calculated
   molecular structure to within 5~\%. The experimental approach allows for the imaging of intrinsic
   molecular dynamics in the molecular frame, albeit this requires more experimental data, which
   should be readily available at upcoming high-repetition-rate facilities.
\end{abstract}
\maketitle

\section{Introduction}
Coherent diffractive imaging has become a widespread tool for a variety of different experiments and
samples, \eg, ranging from the solid state to the gas phase and from small molecules to large
protein crystals. The idea is that the structure of a system, for example, a molecule, protein, or
virus, determines its function. Thus, extracting structural information in the static or
time-dependent domain helps to drastically increase the knowledge of fundamental processes in
nature. The imaging of structure can be performed by the diffraction of electrons or x-rays off the
sample molecules. Electron diffraction has been used for decades to determine the structure of small
gas-phase molecules~\cite{Allen:ActaCryst3:46, Hargittai:GED:1988}, making use of the electrons'
much higher coherent scattering cross section~\cite{Henderson:QRP28:171}. X-rays have a lower
scattering cross section than electrons and hence penetrate more deeply into the sample.
Consequently, x-ray diffraction is often used to image much denser crystalline samples, which, due
to the many identical, oriented molecules, provides a coherent amplification of the signal over the
noise at the Bragg diffraction angles. This led, for instance, to the confirmation of the planar
structure of benzene~\cite{Lonsdale:PRSA123:494}, the structure of
penicillin~\cite{Crowfoot:penicillin:1949}, and the structure of the DNA double
helix~\cite{Watson:Nature171:737}. Today, crystallography is still a very successful approach to
probe the structure of, \eg, proteins, protein complexes, and viruses~\cite{Spence:IUCrJ4:322}.
However, not all molecules can be crystallized and, furthermore, dense crystal packing can constrain
molecular conformations and hamper molecular dynamics.

Diffractive imaging of gas-phase molecules is a highly promising tool to unravel the intrinsic
molecular dynamics of chemical processes on ultrafast timescales~\cite{Zewail:ARPC57:65,
   Neutze:Nature406:752, Filsinger:PCCP13:2076, Barty:ARPC64:415}. Time resolved diffraction studies
of small gas-phase molecules in the picosecond range were first employed by electron diffraction at
the beginning of the 21st century~\cite{Ihee:Science291:458, Sciaini:RPP74:096101} and have been
used ever since with laboratory-based electron sources~\cite{Hensley:PRL109:133202,
   Mueller:JPB48:244001}. Recently, much higher time resolution of $\ordsim100$~fs was achieved by
an accelerator-facility based relativistic electron gun~\cite{Yang:NatComm7:11232,
   Yang:Science361:64}. The development of ultrashort and intense hard x-ray laser pulses generated
by x-ray free-electron lasers (XFELs) has also provided the possibility to image structure as well
as structural changes of small gas-phase molecules through x-ray
diffraction~\cite{Kuepper:PRL112:083002, Stern:FD171:393, Minitti:PRL114:255501} on ultrafast
(femtosecond) timescales.

To retrieve the three-dimensional (3D) diffraction volume of a molecule, which can be inverted to
its 3D structure, knowledge about the relative orientation of the imaged sample(s) with respect to
laboratory fixed axes, \ie, the molecular frame, is highly advantageous or simply necessary when
averaging data over multiple molecules. In a crystal each molecule is aligned with respect to the
crystallographic axes. The crystals usually provide enough scattered photons per XFEL pulse to
determine the orientation of the crystal \textit{a posteriori} and, therefore, the orientation of
each molecule~\cite{Chapman:Nature470:73, Spence:RPP75:102601}. This is not possible for single
small molecules due to the low number of scattered photons per molecule
($\ll1$~photons/molecule/pulse). Instead, access to the molecular frame can be achieved by laser
alignment of a single molecule or a molecular ensemble~\cite{Filsinger:PCCP13:2076,
   Barty:ARPC64:415, Spence:PRL92:198102}. The finitely-sampled diffraction pattern of a perfectly
oriented molecular ensemble is equal to the diffraction pattern of the individual
molecule~\cite{Filsinger:PCCP13:2076},$^,$\footnote{At infinite resolution cross-correlation terms
   between the individual molecules could theoretically be measured.} because the ensemble is
generally lacking translational symmetry -- as opposed to a crystal. This has the added benefit that
the scattering signal may be averaged over many XFEL shots~\cite{Stern:FD171:393}.

Here, we present results on the diffractive imaging of controlled gas-phase \dii (C$_4$H$_2$I$_2$S)
molecules. The data were measured at the coherent x-ray imaging (CXI)
instrument~\cite{Liang:JSR22:514} of the Linac Coherent Light Source (LCLS, experiment LG26, October
2014). The molecules were aligned in all three dimensions by an off-resonant, elliptically
polarized, linearly chirped near-infrared laser pulse at the full XFEL repetition rate of
120~Hz~\cite{Kierspel:JPB48:204002}. The aligned molecular ensembles were probed at a photon energy
of 9.5~keV ($\lambda\approx130~\text{pm}$), enabling the measurement of intramolecular atomic
distances. Approximately 2.2~million individual diffraction patterns of the molecular ensemble have
been integrated, and a two-dimensional (2D) diffraction pattern from an ensemble of the
three-dimensionally (3D) aligned planar molecules was recorded. The molecular diffraction pattern
was compared to the simulated molecular diffraction pattern. The experimental setup was designed to
measure ultrafast molecular dynamics on 3D-aligned molecules. Due to large background scatter and
the correspondingly limited signal-to-noise ratio of the measurement, we were only able to acquire
the diffraction pattern of the static equilibrium structure. The 3D alignment of the molecules was
independently verified by velocity map imaging (VMI)~\cite{Eppink:RSI68:3477} of ionic fragments of
the molecules~\cite{Kierspel:JPB48:204002}.

\section{Experimental setup and procedure}
\label{sec:setup}
\begin{figure}
   \includegraphics[width=\linewidth]{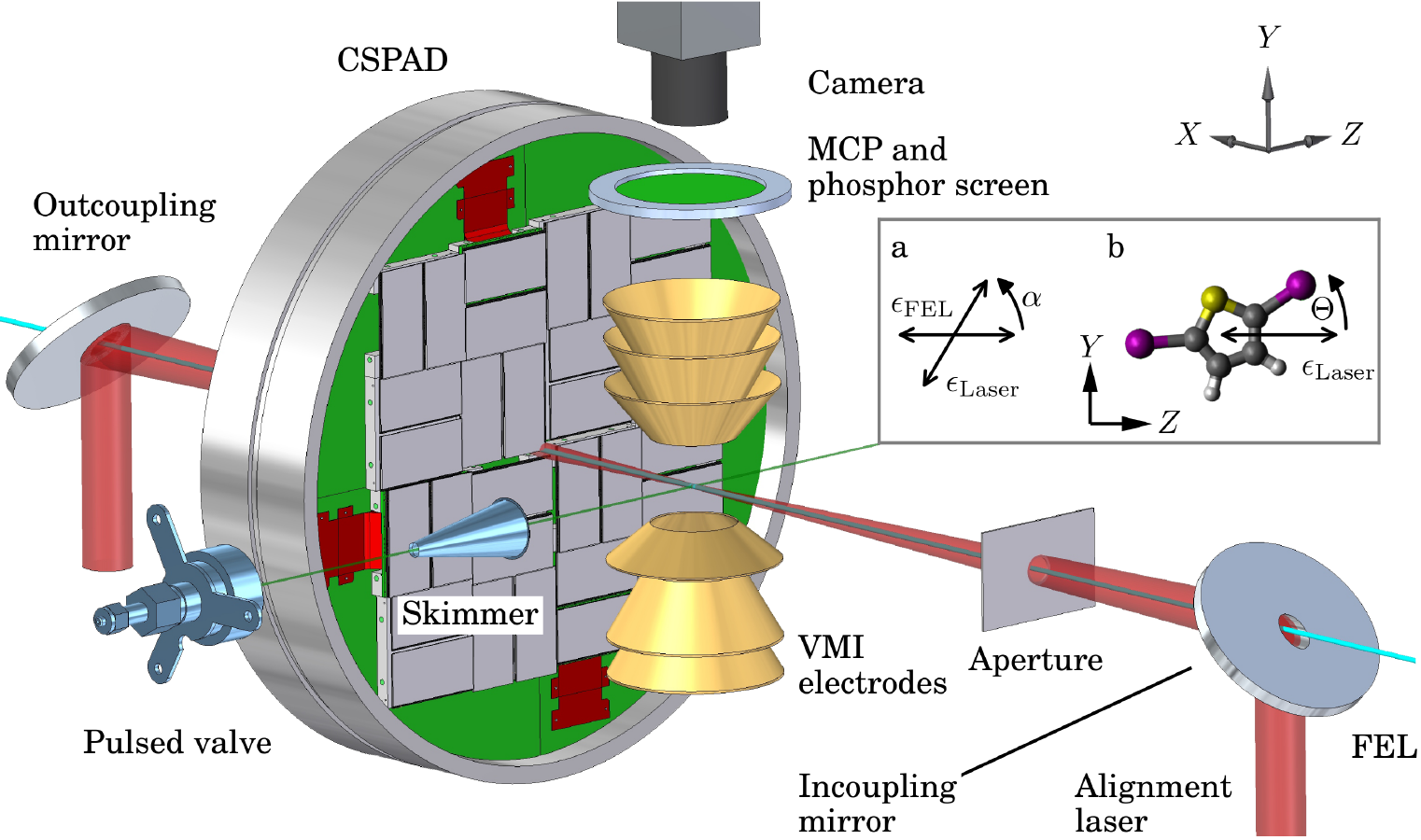}
   \caption{Scheme of the experimental setup showing the pulsed valve, skimmer, and the molecular
      beam axis indicated by the green line, which is crossed by the XFEL (cyan) and the alignment
      laser (red). The ion detection system was used to measure the degree of alignment of the
      molecules, and consists of the VMI electrodes, MCP, phosphor screen, and camera. The CSPAD
      camera was used to record x-ray photons. Two holey in- and outcoupling mirrors were used to
      guide the alignment laser through the vacuum chamber, and to ensure collinear propagation with
      the XFEL. The aperture was used to reduce background from the XFEL beamline on the detector.
      The inset shows in (a) the definition of the angle $\alpha$, which corresponds to the angle
      between the polarizations of the XFEL $\epsilon_{\text{XFEL}}$ and the alignment laser
      $\epsilon_{\text{Laser}}$ and was set to $\degree{0}$ or $\degree{66}$ for the measurement of
      the degree of alignment or the diffraction pattern, respectively. (b) shows the molecular
      structure, and schematically defines the angle $\theta$ between $\epsilon_{\text{Laser}}$, and
      the iodine-iodine (I-I) axis of the molecule.}
   \label{fig:setup}
\end{figure}
The experimental setup is sketched in \autoref{fig:setup}. A detailed description of the molecular
beam parameters as well as the achieved alignment of the molecules is published
elsewhere~\cite{Kierspel:JPB48:204002}. In short, \dii molecules were placed in the sample reservoir
of the pulsed Even-Lavie valve~\cite{Even:JCP112:8068}, which was heated to a temperature of
\celsius{75} at the tip of the valve. The molecules were seeded in 80~bar of helium and
supersonically expanded into vacuum at a repetition rate of 120~Hz, synchronized to the XFEL
repetition rate. A single skimmer (Beam Dynamics, 3~mm diameter) 8~cm downstream of the valve
resulted in a 5.2~mm wide molecular beam (full width at half maximum, FWHM) in the interaction zone.

The molecular-beam pulse duration was on the order of 45~\us (FWHM), which led to a peak density of
$\ordsim1\cdot10^9$ molecules per cm$^{3}$. The molecules were aligned by an in-house
chirped-pulse-amplified Ti:Sapphire (TSL) laser system (Coherent) at full XFEL repetition rate. The
alignment laser, depicted in red in \autoref{fig:setup}, was coupled into the XFEL beam path (cyan)
by a holey incoupling mirror to ensure that both beam paths were collinear. An aperture was placed
in between the incoupling mirror and the interaction zone to reduce scattering reaching the
detectors from sources other than the aligned molecules, such as components in the beamline. The
alignment laser pulses were linearly chirped with a pulse duration of 94~ps (FWHM) and a pulse
energy of 3.3~mJ, focused to 45~\um (FWHM), which resulted in an estimated peak intensity of
$1\cdot10^{12}~\text{W/cm}^2$. The alignment laser was elliptically polarized in the $YZ$ plane with
an aspect ratio of 3:1, and its polarization could be rotated by a $\lambda$/2 waveplate. The XFEL
was linearly polarized along the $Z$ axis and was spatially and temporally placed at the peak
intensity of the alignment laser pulse. It was focused to a spot with a width of 12~\um in the
horizontal and 3~\um in the vertical axis, and had a pulse duration of approximately 70~fs (FWHM) at
photon energy of 9.5~keV, and a pulse energy of approximately 0.64~mJ in the interaction zone,
resulting from $4.2\cdot10^{11}$~photons, a beam line transmission of 80~\%, and a focusing-optics
transmission of 40~\%. The degree of alignment (DOA) of the molecules was probed \emph{via}
ion-momentum imaging perpendicular to the molecular and laser beams in a VMI spectrometer consisting
of the VMI electrodes, microchannel plate (MCP), phosphor screen, a fast high-voltage switch
(Behlke), and a CCD camera (Adimec Opal)~\cite{Kierspel:JPB48:204002}. Diffracted photons were
measured with the Cornell-SLAC Pixel Array Detector (CSPAD)~\cite{Hart:PSPIE8504:85040C} 8~cm
downstream of the interaction zone. The XFEL and the alignment laser were guided through a central
hole of the CSPAD camera. The outcoupling mirror was used to steer the alignment laser outside of
the vacuum chamber.

The molecular DOA was probed by rotating the major axis of the alignment laser polarization
$\epsilon_{\text{Laser}}$ in the $YZ$ plane such that it was parallel to the VMI detector plane,
\ie, $\alpha=\degree{0}$ in \autoref[(inset~a)]{fig:setup}. The most polarizable axis of the \dii
molecules -- an axis parallel to the iodine-iodine (I-I) axis -- aligned along the major axis of the
alignment laser polarization ellipse. The second most polarizable axis aligned along the minor axis
of the alignment laser polarization, leading to a 3D-aligned molecular
ensemble~\cite{Stapelfeldt:RMP75:543}; weak 3D orientation might have been present due to the dc
electric field from the VMI~\cite{Nevo:PCCP11:9912}, but is not of further relevance. The molecules
were Coulomb exploded by the XFEL and VMI spectra of different ionic fragments such as
$\text{I}^{+2}$ or $\text{I}^{+3}$ were recorded~\cite{Kierspel:JPB48:204002}. Due to the high
degree of axial recoil for these ionic fragments they allowed for an accurate determination of the
DOA~\cite{Kierspel:JPB48:204002}, typically quantified by \cost. Here, $\theta$ is defined as the
angle between the major axis of the alignment laser polarization and the axial recoil axis of the
molecule, see \autoref[(inset b)]{fig:setup}. $\theta_\text{2D}$ is the corresponding projected
angle in the $XZ$-plane, which is measured by the VMI spectrometer. \cost ranges from 0.5 to 1 for
an isotropic and a perfectly aligned molecular ensemble, respectively. When measuring diffraction
from the aligned molecular ensemble, we rotated the major axis of the alignment laser polarization
to $\alpha=\degree{66}$ as a compromise between the maximum amount of scattered photons at
$\alpha=\degree{90}$, and the largest observable scattering vector at $\alpha=\degree{45}$. The
molecular DOA was regularly confirmed between diffraction runs by switching between the recording of
diffraction images at $\alpha=\degree{66}$ and VMI images at $\alpha=\degree{0}$.

\section{Simulations}
\label{sec:xraydiff:sim}
The simulations of the diffraction pattern of \dii were carried out using the CMIdiffract code,
which was developed within the CMI group to simulate the diffraction of x-rays or electrons of
gas-phase molecules based on the independent atom model~\cite{Kuepper:PRL112:083002,
   Stern:FD171:393, Stern:thesis:2013, Mueller:thesis:2016, Kierspel:Dissertation:2016,
   Mueller:JPB48:244001}.

The structure of \diis, which was used to calculate the diffraction pattern, was computed with
GAMESS-US~\cite{Gordon:GAMESS:2005} at the MP2/6-311G** level of theory. Parameters such as
molecular beam density, molecular beam width, and the degree of alignment were extracted from the
experiment~\cite{Kierspel:JPB48:204002} and appropriately considered in the simulations, as were
geometric constrains such as the distance from the interaction zone to the CSPAD camera, the size of
the detector, the number of incident photons and their energy as well as polarization. Contributions
from dissociating molecules in the diffraction pattern are estimated to be on the order of
2~\%~\cite{Kuepper:PRL112:083002, Stern:FD171:393} and are neglected in the calculated diffraction
patterns.

\begin{figure}
   \includegraphics[width=\linewidth]{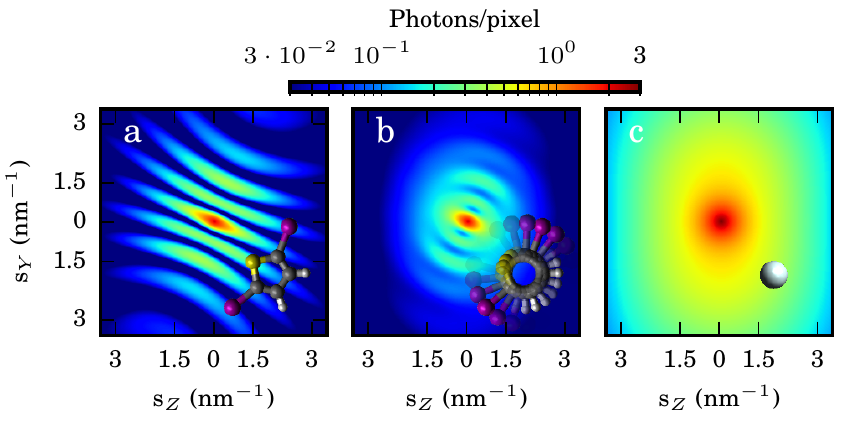}
   \caption{Simulated diffraction pattern of \dii at the location of the CSPAD camera for (a)
      perfectly aligned and (b) strongly aligned molecules as well as (c) for helium. The insets in
      the patterns schematically show the molecular structures adapted to the employed DOA and the
      helium atom.}
   \label{fig:simdegali}
\end{figure}
\autoref[a]{fig:simdegali} shows the simulated diffraction pattern on the detector for a perfectly
aligned molecular ensemble scaled to the number of acquired XFEL pulses for this experiment, \ie,
$\ordsim2.2\times10^{6}$ XFEL pulses, corresponding to 5.2~h of data acquisition at 120~Hz. The
color scale is given by the amount of photons per pixel at a resolution of $1736\times1736$, and the
axes are given as the scattering vector $s$.\footnote{The scattering vector $s$ is defined by
   $s=\text{sin}(\Theta)/\lambda$, with the scattering angle $\Theta$; therefore, 2$\Theta$ is
   defined between the axis of the XFEL and a point on the detector.} In the insets sketches of the
molecular structure and its orientation for one out of two possible orientations for a 3D-aligned
molecule with the given alignment laser polarization, \autoref{sec:setup}, are shown for
illustration purposes; in the calculations the correct probability densities are used.

The diffraction patterns show a ``double-slit like'' interference pattern of the molecule, which is
caused by the significantly larger coherent scattering cross section of the iodines compared to the
other atoms in the molecule~\cite{Berger:XCOM:V15}. The increased bending of the fringes towards
higher $s$ is due to the projection of the Ewald's sphere onto a flat detector surface. The
iodine-sulfur cross correlation is the second strongest contributor to the diffraction pattern.
Every second maximum of the iodine-iodine pattern has contributions from it, since the sulfur is
half way between the iodines nearly on the same internuclear axis.

\autoref[b]{fig:simdegali} shows the simulated diffraction pattern for the same parameters as in
\autoref[a]{fig:simdegali}, but calculated for the experimentally determined degree of
alignment~\cite{Kierspel:JPB48:204002} averaged over the course of the whole data run, $\cost=0.81$.
The inset schematically visualizes the width of the alignment distribution of the molecules.
Compared to perfectly aligned molecules, the contrast of the fringes is reduced and the diffraction
pattern is washed out.

\autoref[c]{fig:simdegali} shows the structureless diffraction pattern for helium atoms for an
estimated helium to molecule ratio of 8000:1 -- corresponding to 10~mbar vapor pressure of the
molecules seeded in 80~bar of helium (\emph{vide supra}). At this ratio the number of scattered
photons from the helium is around 0.5 scattered photons per XFEL pulse, which is 5 times higher than
the signal from the aligned molecules. While this helium background can be strongly reduced using
the electric deflector~\cite{Chang:IRPC34:557, Trippel:RSI89:096110}, this approach was not used
here in favor of a shorter length of the molecular beam path and correspondingly higher densities.

\begin{figure}
   \includegraphics[width=\linewidth]{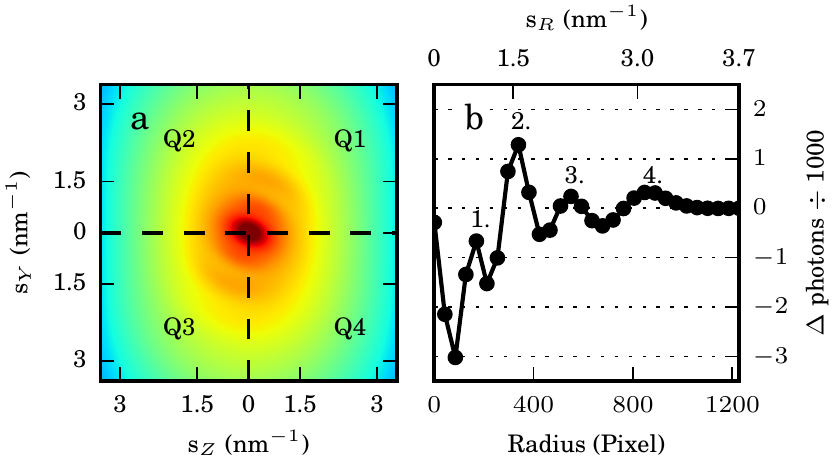}
   \caption{(a) Sum of the calculated diffraction paterns of helium, \autoref[c]{fig:simdegali}, and
      the strongly aligned \dii molecules, \autoref[b]{fig:simdegali}; the color scale is the same
      as in \autoref{fig:simdegali}. (b) Radial differences $\Delta\mathrm{q}$ between the quadrants
      Q1--Q4 extracted from (a). Non-zero values result from correspondingly anisotropic features in
      the diffraction patterns, which originate solely from the aligned molecular diffraction
      pattern; see text for detail. The locations of the first four maxima of the I--I interference
      fringes are highlighted by the numbers 1--4; see text for details.}
   \label{fig:sim:2d_and_1d}
\end{figure}
The expected diffraction pattern, \ie, the sum of \autoref[b,c]{fig:simdegali}, is shown in
\autoref[a]{fig:sim:2d_and_1d} with the same color scale as shown in \autoref{fig:simdegali}. The
contrast of the diffraction pattern is strongly reduced due to the contribution from the
helium-seed-gas scattering, but the general features are still visible. The contribution of the
seed-gas scattering and isotropic background from rest gas in the chamber can be removed as
described in the following procedure: Q1\ldots{}Q4 represent the different quadrants of the
detector, see \autoref[a]{fig:sim:2d_and_1d}. Even with the x-ray-polarization factor included, the
diffraction pattern of atoms or isotropic molecules is symmetric with respect to $Y$ and $Z$, \ie,
$\text{Q1(s}_Y,\text{s}_Z)=\text{Q2(s}_Y,-\text{s}_Z)=\text{Q3(-s}_Y,-\text{s}_Z)=\text{Q4(-s}_Y,\text{s}_Z)$.
Due to the 3D alignment of the molecules at $\alpha\neq{}n\cdot\degree{90},n=0,1,2\ldots$, the
molecular diffraction pattern obeys the symmetries
$\text{Q1(s}_Y,\text{s}_Z)=\text{Q3(-s}_Y,-\text{s}_Z)\neq\text{Q2(s}_Y,-\text{s}_Z)=\text{Q4(-s}_Y,\text{s}_Z)$.
Therefore, the radial distributions $\text{s}_R$ for the quadrants, labeled as
q$_{1}$\ldots{}q$_{4}$, obey the symmetries
$\mathrm{q_{1}(s}_R)=\mathrm{q_{2}(s}_R)=\mathrm{q_{3}(s}_R)=\mathrm{q_{4}(s}_R)$ for the
diffraction off atoms and isotropic molecules, and
$\mathrm{q_{1}(s}_R)=\mathrm{q_{3}(s}_R)\neq\mathrm{q_{2}(s}_R)=\mathrm{q_{4}(s}_R)$ for the aligned
molecules.

Calculating $\Delta\mathrm{q=(q_{1}+q_{3})-(q_{2}+q_{4})}$ for the simulated diffraction pattern
shown in \autoref[a]{fig:sim:2d_and_1d} results in a radial distribution solely dependent on the
summed molecular diffraction patterns, which is shown in \autoref[b]{fig:sim:2d_and_1d}. Here, the
first four maxima of the I--I interference term, highlighted by the numbers 1--4, are visible. The
fringes are clearly visible with a strong contrast over the background. The location of the maxima
along this radial diffraction pattern mainly depend on the molecular structure, whereas their
relative amplitudes are dependent on the DOA and $\alpha$. The spacing of the fringes changes with
the radius due to the projection of the Ewald sphere onto the planar detector.

This data shown in \autoref[b]{fig:sim:2d_and_1d} looks similar to the so-called modified scattering
intensity $sM(s)$, which is frequently used in the data analysis of, \eg, gas-phase
electron-scattering experiments. However, the approach used here intrinsically suppresses isotropic
features in the diffraction pattern, and is, therefore, only applicable to single- or
aligned-molecule ensembles, and not applicable to the diffraction of isotropically oriented
molecules.

\section{Results and discussion}
\begin{figure}
   \includegraphics[width=\linewidth]{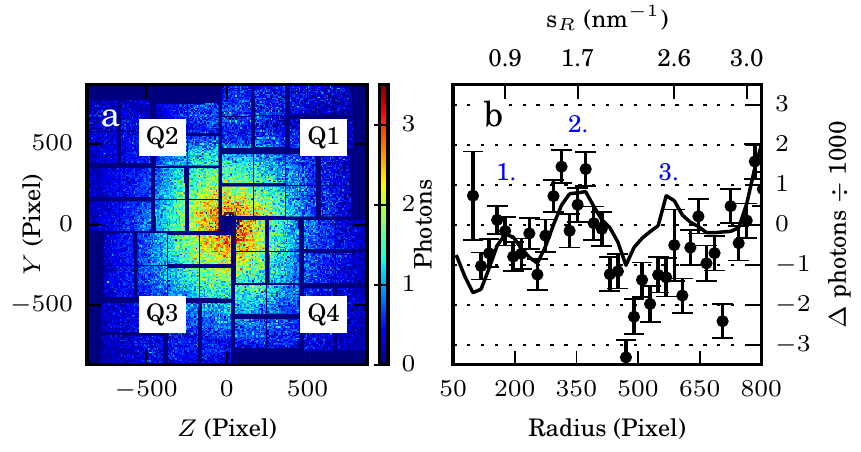}
   \caption{(a) Experimental background-corrected diffraction pattern recorded with the CSPAD
      camera; the horizontal and vertical dark stripes are due to gaps in the detector. Quadrants
      are labeled by Q1--Q4, see text for details. (b) \dqn for the simulated (solid) and measured
      (dots with error bars) diffraction pattern. The locations of the first three maxima of the
      iodine-iodine interference are labeled 1--3.}
   \label{fig:xraydiff:results}
\end{figure}
\autoref[a]{fig:xraydiff:results} shows the measured diffraction pattern for aligned \dii molecules
seeded in helium. For this image $\ordsim2.2\times10^{6}$ individual diffraction images have been
integrated and background corrected to compensate for photons originating from the beamline, which
contributed $\ordsim30~\%$ to the detected number of photons. The background correction was
performed by subtracting averaged images from measurements without molecular beam, \ie, the
molecular beam was either switched off or temporally delayed such that the XFEL pulses missed it.
This resulted in the diffraction pattern shown in \autoref[a]{fig:xraydiff:results}. For
illustration purposes, the recorded diffraction was averaged between neighboring pixels over a
$7\times8$ pixel window, resulting in a 2D diffraction pattern with decreased pixel-based
fluctuations and largely avoided negative intensities resulting from the background correction. The
horizontal and vertical dark stripes in the image are due to gaps in the CSPAD detector, \cf
\autoref{fig:setup}. Q1\ldots{}Q4 label the different quadrants of the detector as in
\autoref[a]{fig:sim:2d_and_1d}.

\autoref[b]{fig:xraydiff:results} shows the radial difference between the quadrants, \dqn for the
simulation (solid line) and the experiment (points), \cf \autoref[b]{fig:sim:2d_and_1d}. Unlike
$\Delta\text{q}$, \dqn contains a radius-dependent correction factor accounting for the lower number
of summed pixels per bin due to the gaps of the detector. The error bars for the experimental data
are given as one standard deviation; we note that they are largely independent of radius due to the
distribution of beamline-scatter background with most intensity on the outer part of the detectors,
which practically cancels the expected noise-distribution from the signal itself. The simulated
diffraction pattern was modified by the gaps of the detector before \dqn was calculated.

The simulations show that the first three maxima of the I--I interference are clearly visible
despite of the gaps; the fourth maximum is already strongly influenced by missing pixels and,
therefore, is not shown any more. For radii $\lesssim100$~pixel ($\text{s}_R<0.04$) the measured
signal strongly deviates from the simulations, with ordinate values higher or lower than the shown
range. This is attributed to stray photons from the direct x-ray beam, which are strongly observable
close to the central hole of the detector.

The first two maxima of the I--I interference pattern -- and hence the first maximum of the I--S
interference -- are matched well by the measurement, including a change of sign around the second
maximum. At higher scattering angles the deviation between measurement and simulations is
increasing. Here, the intensities in the measurement are overall smaller than in the simulation, but
the general trend of an increased signal around the third maximum is comparable. The deviation is
assigned to the small diffraction signal for scattering angles $s_{R}>0.2$, which leads to an larger
influence of the measured background photons.

\begin{figure}
   \includegraphics[width=\linewidth]{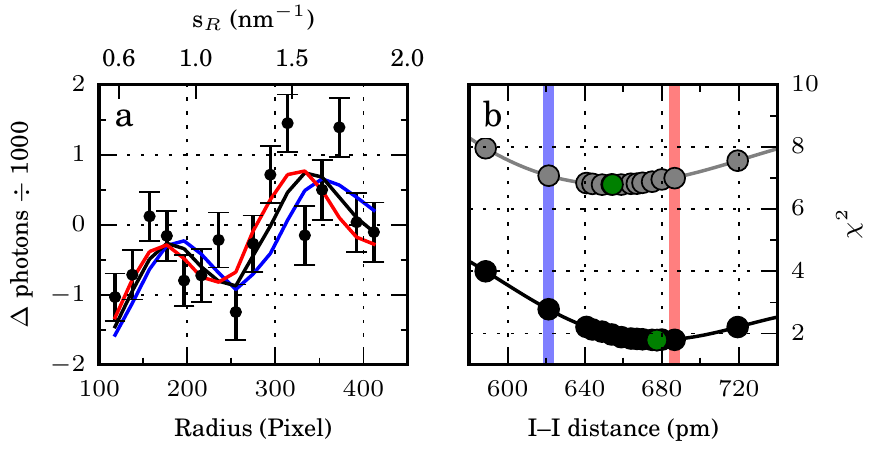}
   \caption{(a) \dqn for the first two maxima of the diffraction pattern. The solid lines show the
      simulated diffraction pattern for the calculated I--I bond distance (black), and a varying
      bond distances in a range +/- 5 \%, \ie, 686.7 and 621.3 pm (red, blue). See
      \autoref{fig:xraydiff:results} for the comparison between experiment and simulation over the
      full radial range. (b) Reduced chi-square value (points) in dependence of different simulated
      I--I bond distances for two scattering ranges $\mathrm{s_{R1}}$ (black, same as (a)), and
      $\mathrm{s_{R2}}$. Polynomial fits (solid) were used to determine the respective minima, which
      are highlighted by the green points. As in (a), the vertical red and blue lines show the +/- 5
      \% range from the calculated bond distance. }
   \label{fig:xraydiff:chi}
\end{figure}
The I--I bond length is reconstructed from the experiment by a comparison to several simulated
molecular diffraction patterns with varying I--I bond distance. In \autoref[a]{fig:xraydiff:chi} the
experimental data are compared to simulated \dqn for three different I--I bond distances, namely the
computed equilibrium distance of 654~pm (black), \emph{vide supra}, and for a variance of $\pm5$~\%
of the I--I bond (red/blue). The I--I distances were varied by symmetrically elongating the iodines
along the connecting line while keeping the rest of the molecular structure unchanged. By focusing
on the scattering range $\mathrm{s_{R,1}}=[0.58,2]~\text{nm}^{-1}$, which contains the first two
maxima of the I--I interference pattern, the simulations already show that changes on the order of
$\pm5$~\% in distance shift the radial maxima inevitably toward higher and lower scattering angles,
respectively.

In order to quantitatively determine the best-fit I--I distance for the experimental data, we
performed a $\chi^2$ analysis of the simulations against the experimental data. The black points in
\autoref[b]{fig:xraydiff:chi} show the reduced $\chi^2$
values~\footnote{$\chi_{\text{Red}}^2=\frac{\sum_{n=1}^{N}(\frac{y_n-f(n)}{\sigma_n})^2}{N-1}$,
   where N is the total number of considered radial bins, n is $x-$value corresponding to a radial
   bin, $y$ is the corresponding experimental determined value with a standard deviation $\sigma$,
   and $f(n)$ is the corresponding value from the simulated diffraction pattern.} for different I--I
distances for the scattering range $\mathrm{s_{R1}}$; the gray points show the same analysis for a
scattering range of $\mathrm{s_{R,2}}=[0.58, 3.13]~\mathrm{nm^{-1}}$, which includes the third
maxima of the I--I interference pattern. The corresponding solid lines show polynomial fits to the
$\chi^2$-values. The fits provide an optimized bond distance of $677$~pm and $654$~pm for the
scattering ranges $\mathrm{s_{R,1}}$ and $\mathrm{s_{R,2}}$ with corresponding $\chi^2$ values of
$1.8$ and $6.8$, respectively. Both minima are highlighted by an additional green point. The
retrieved distances are in very good agreement with the quantum-chemistry distance of 654~pm, and
are clearly within $\pm5$~\% of the calculated I--I distance, indicated by the vertical red (blue)
line.

\section{Summary and outlook}
We presented experimental results on the diffractive imaging of controlled gas-phase molecules. The
molecules were strongly aligned by an in-house TSL, which allowed the measurement of diffraction
patterns at the full LCLS repetition rate~\cite{Kierspel:JPB48:204002} of 120~Hz. The aligned
molecules were probed with hard x-ray photons at a photon energy of 9.5~keV. The molecular
diffraction pattern of a 3D-aligned molecular ensemble was successfully extracted, as confirmed by
utilizing the different symmetries in the diffraction pattern of the aligned molecules and the seed
gas. The extracted iodine-iodine distance was in agreement with the calculated molecular structure
to within a few percent.

We note that this iodine-iodine distance in the static molecule could also have been measured using
gas electron diffraction or x-ray scattering off a dense isotropic gas. However, the current
experimental demonstration of this measurement using aligned molecules and femtosecond pulses of
hard x-rays provides crucial steps toward the coherent diffractive imaging of ultrafast molecular
dynamics at the atomic scale, the long-sought after ``molecular movie'': Conceptionally, the aligned
molecules~\cite{Stapelfeldt:RMP75:543, Holmegaard:NatPhys6:428, Kierspel:JPB48:204002,
   Trippel:JCP148:101103, Owens:JPCL9:4206} enable the recording of the three-dimensional coherent
diffraction image of the molecule instead of a one-dimensional radial scattering distribution, \ie,
in principle, it provides information on the three-dimensional structure of molecules instead of the
pair-distribution functions obtained from standard electron or x-ray scattering. The femtosecond
x-ray pulses provide the means to follow ultrafast dynamics with atomic resolution.

In a previous experiment~\cite{Kuepper:PRL112:083002}, 1D aligned molecules were probed at a much
lower photon energy (2~keV, $\lambda_\text{{XFEL}}=~600~\text{pm}$), which led to a resolvable
structure in the order of the size of the molecule. The larger photon energy in the current
experiment allowed for the quite precise measurement of an intramolecular atomic distance, albeit so
far with a decreased signal-to-noise ratio (SNR) due to the lower coherent scattering cross section,
higher incoherent scattering cross sections, and a reduced photon flux from the LCLS facility. The
SNR, largely limited by large background contributions from the beamline and the last x-ray aperture
before the interaction region, was really the limiting factor in this experiment as, for example,
the obtained 2D diffraction pattern of the static structure of \dii shown in
\autoref[a]{fig:xraydiff:results} is very noisy. A comparison between experiment and simulation was
only useful by improving the SNR in the data analysis by summing up neighboring pixels, and
calculating differential radial plots as shown in \autoref[b]{fig:xraydiff:results}.

Based on our simulations we estimate that for this measurement the difference between the
diffraction pattern of 1D and 3D aligned \dii molecules would be negligible. However, we further
estimate that the number of acquired individual diffraction patterns was sufficient to distinguish
between 1D and 3D-aligned molecular ensembles if the molecular degree of alignment was close to 1,
hinting at the possibility to determine the complete molecular structure.

The experimental setup was technically capable of investigating ultrafast molecular dynamics: The
setup provided a collinearly aligned femtosecond laser pulse, which was powerful enough to
dissociate the aligned molecules~\cite{Kierspel:JPB48:204002}. But the measurement of molecular
dynamics requires a higher number of scattered photons or an improved SNR.

Experimentally, the SNR can be improved by reducing the number of background photons on the detector
via, for example, optimized x-ray apertures, or by the implementation of the electric
deflector~\cite{Chang:IRPC34:557, Trippel:RSI89:096110} into the experimental setup. The deflector
is placed between the valve and the interaction zone, and allows to spatially separate polar
molecules from the seeding gas. This technique was applied once for the diffractive imaging of
controlled molecules~\cite{Kuepper:PRL112:083002, Stern:FD171:393}, but was not applied here due to
the corresponding longer distance from valve to interaction point, resulting in a lower density of
the molecular beam.

The repetition rates of the recently launched European XFEL or the upcoming LCLS~II are a few
hundred to a few thousand times higher than available here and these facilities will provide a
near-infrared laser synchronized with the XFEL, which will align molecules at very high repetition
rates~\cite{Trippel:MP111:1738, Kierspel:JPB48:204002}; alternatively, continuous-wave alignment
could be exploited~\cite{Deppe:OE23:28491}. If the molecular alignment is achieved at the full
repetition rates of these upcoming facilities, the presented experiment results can be measured
within two minutes at the European XFEL or a few seconds at the LCLS~II. Such experimental
parameters provide a feasible start for the recording of ultrafast molecular dynamics of small
3D-aligned gas-phase molecules~\cite{Barty:ARPC64:415}, or to image small biomolecules without heavy
atoms in the molecular frame.

\begin{acknowledgments}
   This work has been supported by the Deutsche Forschungsgemeinschaft through the Clusters of
   Excellence ``Center for Ultrafast Imaging'' (CUI, EXC~1074, ID~194651731) and ``Advanced Imaging
   of Matter'' (AIM, EXC~2056, ID~390715994), the Helmholtz Association through the Virtual
   Institute 419 ``Dynamic Pathways in Multidimensional Landscapes'' and the ``Initiative and
   Networking Fund'', the European Union's Horizon 2020 research and innovation program under the
   Marie Sk{\l}odowska-Curie Grant Agreement MEDEA (641789), and the European Research Council under
   the European Union's Seventh Framework Programme (FP7/2007-2013) through the Consolidator Grant
   COMOTION (ERC-Küpper-614507).

   Use of the Linac Coherent Light Source (LCLS), SLAC National Accelerator Laboratory, is supported
   by the U.S.\ Department of Energy, Office of Science, Office of Basic Energy Sciences
   (DE-AC02-76SF00515). D.R.\ and A.R.\ acknowledge support from the Chemical Sciences, Geosciences,
   and Biosciences Division, Office of Basic Energy Sciences, Office of Science, U.S.\ Department of
   Energy (DE-FG02-86ER13491).
\end{acknowledgments}

\bibliography{string,cmi}%
\end{document}